\documentstyle [11pt,psfig]{article}
\begin{document}
\begin{center}{
{\bf  Empirical Emission Functions  for
LPM Suppression of Photon Emission from Quark-Gluon Plasma }
\\
{S. V. S. Sastry}\\
{ Nuclear Physics Division, Bhabha Atomic Research Centre,
Trombay, Mumbai 400 085, India}
\vspace {-.4cm}
}\end{center}

\par
\vspace {1.0cm}
\centerline{Abstract}
\par

\noindent
The  LPM  suppression  of  photon  emission rates from the quark gluon
plasma have been studied  at  different  physical  conditions  of  the
plasma  given  by  temperature  and  chemical potentials. The integral
equation    for    the    transverse    vector     function     (${\bf
\tilde{f}(\tilde{p}_\perp   )}$)  consisting  of  multiple  scattering
effects is solved by the  variational  method  as  well  as  the  self
consistent     iterations     method    for    the    parameter    set
\{$p_\parallel,k,\kappa,T$\},  for  bremsstrahlung   and   $\bf   aws$
processes.      The     peak     positions     of     these     (${\bf
\tilde{f}(\tilde{p}_\perp   )}$)    distributions    for    the    set
\{$p_\parallel,k,\kappa,T$\} depend only on the dynamical variable $x$
defined                                                            by,
$x=\frac{T}{\kappa_0}\left|\frac{1}{p_\parallel}-\frac{1}{p_\parallel+k}\right|$.
The integration over these distributions multiplied by $x^2$ factor is
also shown to depend on this variable $x$, leading to  a  {\it  unique
global  function} $g(x)$ for all temperatures and chemical potentials.
Empirical fits to this {\it dimensionless emission function},  $g(x)$,
are   obtained.   The  photon  emission  rate  calculations  with  LPM
suppression effects reduce to one dimensional integrals  that  involve
folding  over  the  empirical  $g(x)$  function with appropriate quark
distribution functions and the kinematic factors. Using this approach,
the suppression factors for both bremsstrahlung  and  $\bf  aws$  have
been  estimated  for  various  chemical  potentials  and compared with
results of rigorous calculations using variational method. It has been
found that the $k/T$ is good scale for suppression  factors  only  for
zero  density case. At finite density the bremsstrahlung and $\bf aws$
suppression factors versus $k/T$ are temperature dependent.

\vspace {1.cm}
\par
\noindent
Photons  production  is  known  to be an important signal of the quark
gluon plasma formation (QGP) expected in the  relativistic  heavy  ion
collisions.  Photons  are  emitted  at  various  stages  during plasma
evolution and for  an  overview  one  may  see  \cite{peitz}  and  the
references  therein.  The  processes of bremsstrahlung and ${\bf aws}$
arise at effective two loop level and contribute at the leading  order
$O(\alpha\alpha_s)$   owing  to  the  collinear  singularity  that  is
regularized by the effective thermal masses \cite{auren1} . The higher
order multiple scatterings may also contribute at the  same  order  as
the         one         and         two         loop         processes
\cite{auren2,auren3,auren4,arnold1,arnold2,arnold3}. Further, multiple
soft scatterings of the fermion  during  photon  emission  reduce  the
emitted  photon  coherence lengths, known as Landau-Pomeranchuk-Migdal
(LPM) effect. The photon emission rates are suppressed  owing  to  the
LPM  effects \cite{auren3,arnold1,arnold2}, especially as shown by the
suppression factors in Fig. 7. of \cite{arnold2}. It  has  been  shown
that  the  bremsstrahlung  radiation  is strongly suppressed to almost
$~20\%$ at very low $k/T$ values, whereas  the  photon  emission  from
$\bf  aws$ falls strongly for higher $k/T$ values \cite{arnold2}. Thus
the LPM suppression affects  opposite  ends  of  the  photon  emission
spectrum for bremsstrahlung and $\bf aws$ processes.

\par
\noindent
The  photon  production  rates  from  bremsstrahlung and the $\bf aws$
processes have been estimated in \cite{auren1}, in terms of simple one
dimensional  momentum  integrals  and  the  dimensionless   quantities
$J_T,J_L$.  The  $J_T$  and $J_L$ weakly depend on the thermal masses,
only through $m_g^2/m_\infty^2$ \cite{auren1} and  therefore  are  not
very  sensitive to temperature and the chemical potentials. The photon
(energy $k_0$) differential emission rate per unit volume without  and
with LPM effects are given by ${\cal R}^0$ and $\cal R$ respectively.

\begin{equation}
{{\cal R}^0}_{b,a} =
{\cal C}_k  ~ \int dp\left[p^2+(p+k)^2\right] \left[n_B(k_0)(n_f(p)-n_f(p+k))\right]~(J_T-J_L)
\end{equation}
\begin{equation}
{\cal R}_{b,a}= \frac{80\pi T^3\alpha\alpha_s}{(2\pi)^3 9\kappa}\int dp_\parallel
\left[\frac{p_\parallel^2+(p_\parallel+k)^2)}{p_\parallel^2(p_\parallel+k)^2)}\right]\left[n_f(k+p_\parallel)(1-n_f(p_\parallel))\right]
~\int \frac{\bf d^2p_\perp}{(2\pi)^2}  2{\bf \tilde{p}_\perp \cdot\Re\tilde{f}(\tilde{p}_\perp)}
\end{equation}

\par
\noindent
In  the  above  ~~${\cal C}_k=\frac{40\alpha\alpha_sT}{9\pi^4k^2}$ and
$\kappa=m_\infty  ^2/m_D^2  ~~(\kappa_0=\frac{1}{4}  ~~~\mbox{for}~~  \mu=0~~~
\mbox{case})$. The subscripts $(b,a)$ are for bremsstrahlung and ${\bf
aws}$  with  different  kinematic domains and appropriate distribution
functions. $\kappa $ depends on the physical condition of  the  plasma
such as temperature, baryon density, quark and gluon fugacities and is
determined  by the thermal mass ratios. In this work we consider a two
flavor three color case with $\alpha_s=0.2$.

\noindent
$\Re{\bf  \tilde{f}(\tilde{p}_\perp )}$ in Eq. 2 is the real part of a
transverse vector function which consists of the LPM  effects  due  to
multiple  scatterings. This can be taken as transverse momentum vector
${\bf (\tilde{p}_\perp )}$  times  a  scalar  function  of  transverse
momentum  ${\tilde{p}_\perp  }$.  The  sign  ~$\tilde{}$~  denotes the
dimensionless quantities in units of Debye mass $m_D$  as  defined  in
\cite{arnold2}.             The             function             ${\bf
\tilde{p}_\perp\cdot\Re\tilde{f}(\tilde{p}_\perp )}$ is determined  by
the collision kernels ($\tilde{C}({\bf \tilde{q}_\perp})$) in terms of
the  following integral (AMY) equation as derived by Arnold, Moore and
Yaffe (for details see \cite{arnold2}).

\begin{equation}
2{\bf \tilde{p}_\perp}=i\delta \tilde{E}({\bf \tilde{p}_\perp},p_\parallel,k)
{\bf \tilde{f}}({\bf \tilde{p}_\perp},p_\parallel,k) +
\int\frac{d^2{\bf \tilde{q}_\perp}}{(2\pi)^2}
\left[{\bf \tilde{f}}({\bf \tilde{p}_\perp},p_\parallel,k)
-{\bf \tilde{f}}({\bf \tilde{p}_\perp+\tilde{q}_\perp},p_\parallel,k)\right]
\tilde{C}({\bf \tilde{q}_\perp})
\end{equation}

\begin{equation}
\tilde{C}({\bf \tilde{q}_\perp})=\kappa\int d\tilde{q}_\parallel d\tilde{q}^0
\delta(\tilde{q}^0-\tilde{q}_\parallel) \frac{1}{\tilde{q}}\left[
\frac{2}{|\tilde{q}^2-\tilde{\Pi}_L(\tilde{q}^0,\tilde{q})|^2}+
\frac{\left(1-(\tilde{q}^0/\tilde{q})^2\right)^2}
{|(\tilde{q}^0)^2-\tilde{q}^2-\tilde{\Pi}_T(\tilde{q}^0,\tilde{q})|^2}\right]
\end{equation}
\begin{equation}
\delta \tilde{E}({\bf \tilde{p}_\perp},p_\parallel,k)=
\frac{kT}{2p_\parallel(k+p_\parallel)}\left[\tilde{p}_\perp^2+\kappa \right]
\end{equation}

\noindent
Aurenche,  Gelis  and  Zaraket  obtained  an  analytical  form for the
collision kernel in Eq. 44 of \cite{auren4} by establishing sum  rules
for  the  thermal gluon spectral functions. We used this AGZ kernel in
the present work.\footnote{ My earlier reported anomaly of kernel near
$\tilde{p}_\perp\sim 0$ \cite{svs1} was due to a numerical problem. It
was pointed out by Dr.  Francois  Gelis  together  with  the  possible
source  of  error.  The  error  was  corrected and the calculations of
\cite{svs1} were repeated with exact AGZ kernel and the  results  were
not  much  affected. I thank Dr. Gelis for this important correction.}
We solved the integral  equation  for  the  ${\bf
\tilde{p}_\perp\cdot\Re\tilde{f}(\tilde{p}_\perp  )}$  for  the set of
\{$p_\parallel,k,\kappa,T$\} values using both variational method  and
some  test  cases  by  self  consistent  iterations. The present study
includes nine different chemical potential values from 0-2GeV in steps
0f 0.25GeV representing baryon density and for each  density  case  at
five  temperatures  of  T=0.25,  0.35,  0.45, 0.55 GeV. Additional two
cases of quark and gluon  fugacities  representing  unsaturated  phase
space  at  T=0.55GeV  were  studied. The photon energy was binned into
twenty five values for $k/T=0.0-20.0$ whereas the $p_\parallel$ is the
integration variable. As mentioned  in  \cite{svs1},  the  variational
parameter  $A_v$  was  not  optimised. This is taken from an empirical
expression given below as predicted  by  results  of  iterations.  The
$\beta$   and   $C_t$  parameters  were  found  to  vary  weakly  with
temperature. As shown in \cite{svs1} the $A_v$ values are taken to  be
independent  of  chemical  potentials, together with $\beta=-0.62$ and
$C_t=1.0$ for both bremsstrahlung and $\bf aws$ processes.

\begin{equation}
A_v(p_\parallel ,k) = C_t\left(\frac{T}{\kappa_0}\left|\frac{1}{p_\parallel}-\frac{1}{p_\parallel+k}\right|\right)^\beta
\end{equation}
\vspace {-.5cm}

\par
\noindent
The  results of rigorous variational calculations for all the cases of
the  set  of  \{$p_\parallel,k,\kappa,T$\}  have  been  analyzed   for
empirical   understanding.   It  has  been  observed  that  the  ${\bf
\tilde{f}(\tilde{p}_\perp )}$ distributions and therefore  their  peak
positions are not very sensitive to the chemical potentials. It should
be  noted  that  even  for  the  cases  when  the  $A_v$ values of the
variational parameter deviate from the exact peak  position  of  these
distributions, the variational method with large trial set still gives
correct  results.  Peak  search  has  been performed and peak position
values of these distributions are obtained. Figure 1  shows  the  peak
position           values           of           these           ${\bf
\tilde{p}_\perp\cdot\Re\tilde{f}(\tilde{p}_\perp   )}$   distributions
plotted  as  a function photon energies for all $p_\parallel$ for each
$k$. The symbols (b,a) in the curve  labels  in  figure  are  for  the
bremsstrahlung  and  $\bf  aws$  processes  and the temperature values
shown in figure labels. It can be seen that these peak positions  vary
strongly with both $p_\parallel$ and $k$ independently and are process
dependent.

\noindent
It  is interesting to note that the variational parameter depends only
on               the                dimensionless                scale
$x=\frac{T}{\kappa_0}\left|\frac{1}{p_\parallel}-\frac{1}{p_\parallel+k}\right|$.
This is perhaps expected as the AMY equation depends only on the scale
$x$  rather  than  $p_\parallel$  and  $k$ independently. Therefore in
Figure 2, we plotted the peak position  values  corresponding  to  all
cases  in  Fig.  1  versus  the new scale $x$. It can be seen that the
strong $p_\parallel$ and $k$  dependence  has  almost  merged  into  a
single  curve,  establishing  that  $x$ is the only relevant dynamical
scale. It suggests  that  it  is  sufficient  to  study  the  integral
equation     for     the     set    $\{x,T\}$    rather    than    for
\{$p_\parallel,k,\kappa,T$\}.  The  agreement  of   the   (variational
method)  data  in  Fig.  2  can  be  improved  by suitable temperature
dependent factors. In the following, we study the new dynamical  scale
in  more  detail.  For  this  purpose  we  rewrite  the  Eq.  2  by  a
rearrangement of terms as given below.

\begin{equation}
{\cal R}_{b,a}= {\cal C}_k \int dp_\parallel
\left[p_\parallel^2+(p_\parallel+k)^2)\right]\left[n_f(k+p_\parallel )(1-n_f(p_\parallel ))\right]
~C_g~ g(p_\parallel,k,\kappa,T)
\end{equation}

\begin{equation}
 g(p_\parallel,k,\kappa,T)=\left(\frac{\kappa_0}{\kappa}\right)^{0.40}
\left[\frac{kT/\kappa_0}{p_\parallel (p_\parallel+k)}\right]^2~
 \int \frac{{\bf d^2p_\perp}}{(2\pi)^2} 2{\bf \tilde{p}_\perp \cdot\Re\tilde{f}(\tilde{p}_\perp)}
\end{equation}
\begin{equation}
=g(p_\parallel,k,\kappa,T)=\left(\frac{\kappa_0}{\kappa}\right)^{0.40} x^2 \int \frac{{\bf d^2p_\perp}}{(2\pi)^2} 2{\bf \tilde{p}_\perp \cdot\Re\tilde{f}(\tilde{p}_\perp)}
\end{equation}

$$C_g=\frac{\kappa}{T}\left(\frac{\kappa_0}{\kappa}\right)^{-0.40}  \nonumber $$

\par
\noindent
In  these  equations, the multiplicative function ${\cal C}_k$ is same
as the factor appearing in Eq. 1. $C_g$ is the density and temperature
dependent coefficient. The  new  function  $g(p_\parallel,k,\kappa,T)$
contains  the  LPM  suppression effects and any non trivial $\kappa,T$
dependence. Using the results from the  variational  method,  we  have
calculated     the    $g(p_\parallel,k,\kappa,T)$    for    the    set
\{$p_\parallel,k,\kappa,T$\} values. These  values  at  two  different
temperature  are  shown  in  Figure  3  as  a  function of $k$ for all
$p_\parallel$ values. As mentioned in labels in figure , the  data  is
shown for both bremsstrahlung and $\bf aws$ processes at two values of
chemical  potentials.  As  shown  in Fig. 3, the $p_\parallel$ and $k$
dependence is very strong.  The  different  colours  in  Fig.  3  show
different baryon densities, temperatures and processes as mentioned in
the  figure.  Further, there is only a weak dependence on $\kappa$ for
all the cases of baryon densities and fugacities used in  the  present
work.  We  re-plot  in  Figure  4  these  results versus the dynamical
variable $x$. The data of various cases of Fig. 3 merge into a  single
curve  similar to the Fig. 2. Therefore, the function $g$ also depends
only      on      the      variable       $x$,       $\it       i.e.,$
$g(p_\parallel,k,\kappa,T)=g(x)$.   It   should  be  noted  that  this
apparent single curve contains all the data for two temperatures,  two
baryon density values and two processes. Due to merging of all data in
single curve, the other color symbols are overwritten and therefore are
not  visible.  We  parameterized  the curves in Fig. 4 by fitting with
polynomials ($g=\sum_n a_nx^n$) in the range of $x=0.30-2.0$ and by  a
power law ($g=ax^b$) beyond as given by,

$$  g(x)= 2.6709 x^{0.4535460}   ~\mbox{for}~~ x\le 0.20  \nonumber $$
$$  g(x)=\sum_{n=0}^4 a_n x^n ~\mbox{for}~~ 0.2\le x\le 3.0 \nonumber $$
$$  g(x)=\sum_{n=0}^5 b_n x^n   ~\mbox{for}~~ 3.0\le x\le 45.0 \nonumber $$
$$  g(x)=3.20816   ~\mbox{for}~~ x\ge 45.0   \nonumber $$

\noindent
In  above,$a_0=0.798609,  a_1=2.44683,  a_2=-1.61357$,  $a_3=0.551979,
a_4=-0.0720306  $  and  $b_0=2.31193,  b_1=0.18221,   b_2=-0.0148997$,
$b_3=0.584153E-3, b_4=-1.09125E-5,b_5=7.79482E-8$.

\noindent
The purpose of rewriting Eq. 2 in the form of Eq. 7 is for the following
reasons.\\
(1) the $ \int \frac{{\bf d^2p_\perp}}{(2\pi)^2} 2{\bf \tilde{p}_\perp
\cdot\Re\tilde{f}(\tilde{p}_\perp)}$  of  Eq. 2  above  is  very  large
extending over six orders of magnitude, whereas the $g(x)$ obtained by
multiplying this with $x^2$  and  the  density  dependent  factors  is
reduced to just two orders of magnitude as shown on Fig. 4.\\
(2)  It  separates  the  $x$  dependent  terms from the other terms in
emission rate equations.\\
(3)   It  is  made  to  resemble  Eq. 1.  This  implies  that  the  new
dimensionless emission  function  $g(x)$  replaces  the  dimensionless
$J_T-J_L$  term  of Eq. 1 in the presence of LPM effects. This emission
function describes the photon emission of energy $k$, from  quarks  of
momentum  component  $p_\parallel$  and the physical conditions of the
plasma, through a dimensionless dynamical variable $x$.\\
(4)  By  empirical  fits,  the  emission  rates  are  reduced  to  one
dimensional integrals similar to Eq. 1. Further, as  discussed  before,
the  full  distributions are not very sensitive to chemical potentials
and  temperatures.  Therefore,  many  of   the   properties   of   the
dimensionless quantities $J_T-J_L$ may also be satisfied by the $g(x)$
function.\footnote{This work was completed about a month back. However
just few hours before putting on net, I noticed a paper on LPM effects
in di-lepton sector by different method by Aurenche, Gelis, Moore and
Zaraket. This will be studied later.}

\par
\noindent
The  photon emission rates have been calculated for bremsstrahlung and
$\bf aws$ using the empirical $g(x)$ functions for various  values  of
$\mu$.  denoted  by ${\cal R}_{b},{\cal R}_{a}$ in Eq. 7. The emission
rates for effective two loop processes without LPM effects  have  also
been  calculated  using  Eq.  1.  As given in \cite{svs1} the relative
(normalized) suppression factors are given by,

\begin{equation}
S_b(k,\mu )= \left(\frac{{\cal{R}}_b}{{\cal{R}}^0_b}\right)_k \frac{f_b(\mu =0)}{f_b(\mu)}
~~~~\mbox{with}~~~~f_b(\mu)= \left(\frac{{\cal{R}}_b}{{\cal{R}}^0_b}\right)_{k=20T}
\end{equation}
\begin{equation}
S_a(k,\mu )= \left(\frac{{\cal{R}}_a}{{\cal{R}}^0_a}\right)_k \frac{f_a(\mu =0)}{f_a(\mu)}
~~~~\mbox{with}~~~~f_a= \left(\frac{{\cal{R}}_a}{{\cal{R}}^0_a}\right)_{k=0.2T}
\end{equation}

\vspace {-.2cm}

\noindent
These relative suppression factors are shown in Figure 5 for different
($\mu$)  quark  chemical  potentials  for  bremsstrahlung radiation at
$T=0.25$GeV. Figure 6 shows relative suppression factors for $\bf aws$
process. The relative suppression factors obtained from  the  rigorous
calculations  using  variational method with eight trial functions are
also  shown  in  Figs.(5,6)  by  symbols.  Figure  7  shows   relative
suppression  factors for these two processes at $T=0.55$GeV. It can be
seen from Figs.(5-7) that the $k/T$ of the x-axis is a good scale only
for zero density case, ~{\it i.e.,}~ for  different  temperatures  the
zero  density  suppression  curves  are  the  same.  However at finite
density, the relative suppression factors depend on  temperature.  For
example  one may see the $\mu_q=2.0$GeV curve in Figs.(5,7). Using the
present empirical method, we calculated the  bremsstrahlung  and  $\bf
aws$  suppression  factors  for  the case of fugacities for quarks and
gluons  representing  unsaturated  phase   space.   These   fugacities
correspond to extreme cases of gluon dominated ($\lambda
_g$=0.10,$\lambda_g$/$\lambda_q$=10.0)     or      quark      dominated
($\lambda_g$=0.01,$\lambda_g$/$\lambda_q$=0.1)plasma. These results are
in agreement with results from variational calculations for this case.

%\vspace {.5cm}
\newpage
\par
\noindent
{\bf Conclusion}

\noindent
The  photon  emission  rates  from  the  quark  gluon plasma have been
studied considering LPM suppression effects  at  various  temperatures
and chemical potentials which represent the physical conditions of the
plasma.  Self-consistent  iterations method and the variational method
have been used to solve  the  AMY  integral  equation  for  the  ${\bf
\Re\tilde{f}}({\bf    \tilde{p}_\perp})$   distributions.   The   peak
positions of these distributions from the variational method have been
fitted by an empirical expression for various  parameter  sets.  These
peak  positions  are observed to be rather insensitive to the chemical
potentials at all the temperatures considered. It is  shown  that  the
peak positions depend only on a new dynamical variable $x$ rather than
$p_\parallel$   and   $k$   independently.   Further,  the  integrated
distributions multiplied by $x^2$ are shown  to  depend  only  on  the
variable  $x$  given  by $g(x)$. This establishes that $x$ is the only
relevant dynamical variable for photon emission from QGP for these two
processes. Empirical fits  to  the  new  $g(x)$  functions  have  been
obtained.   Using   the   empirical  $g(x)$  functions,  the  relative
suppression  factors  for  bremsstrahlung  and  $\bf  aws$  have  been
estimated  as  a  function  of  photon  energy  for  various  chemical
potentials. These suppression factors are shown to agree well with the
rigorous results using variational method for all the  cases  studied.
At   finite  density  these  bremsstrahlung  and  $\bf  aws$  relative
suppression factors versus $k/T$ are temperature dependent.

%%%%%%--------------------------------------------------------------------
\par
\noindent
{\bf Acknowledgements}

\noindent
I  acknowledge  the  fruitful  discussions  with  Drs. S. Kailas, A.K.
Mohanty and A. Navin. My sincere  thanks  to  the  staff  of  computer
division  for computing facilities and co-operation during this study.
My thanks  to  Dr.  D.C.  Biswas  for  making  his  personal  computer
available to me during this study.

%%%%%%--------------------------------------------------------------------

%%%%%%--------------------------------------------------------------------
%%%------------peak positions of f(p) distributions-----------------%%%%%%%%
\begin{figure}[!ht]
\begin{center}{
\hspace{-.5cm}
\begin{minipage}{16.cm}
\psfig{figure=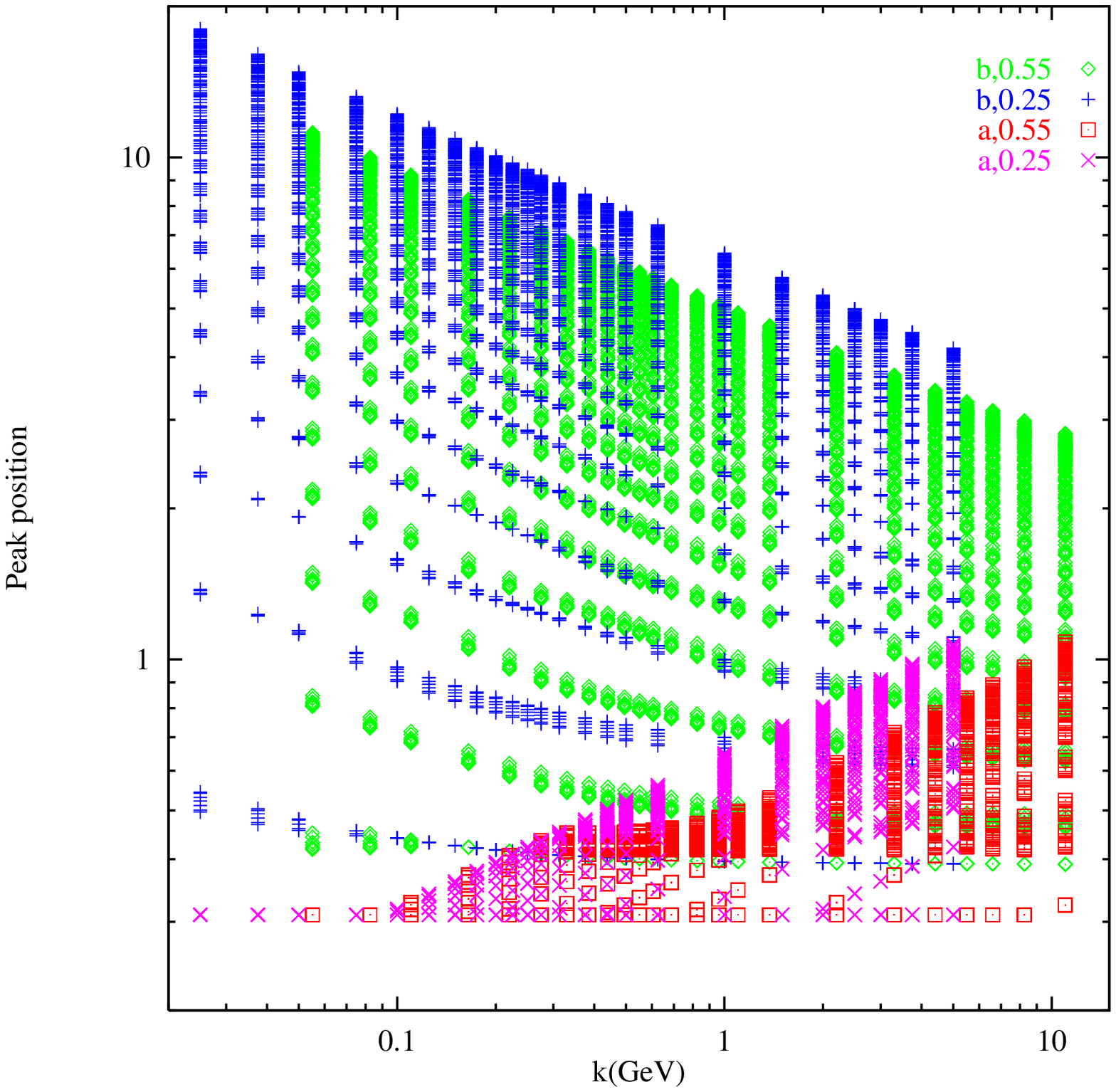,height=18.cm,width=16.0cm}
\vspace{-.50cm}
\caption{  The  peak  positions of the $2{\bf \tilde{p}_\perp}\cdot\Re
{\bf   \tilde{f}}({\bf   \tilde{p}_\perp   })$   distributions    from
variational  method  for  bremsstrahlung  and  ${\bf  aws }$ processes
versus photon energy. The data is shown for two temperatures. For each
process at a given temperature, the data contains
(7x24 points) seven baryon density
values and twenty four $p_\parallel$ values all  in  same  symbol  and
color.}
\end{minipage}
}\end{center}
%\vspace {-0.25cm}
\end{figure}
%\newpage
\begin{figure}[!ht]
\begin{center}{
\hspace{-2.5cm}
\begin{minipage}{16.cm}
%%%%%%--------------------------------------------------------------------
\psfig{figure=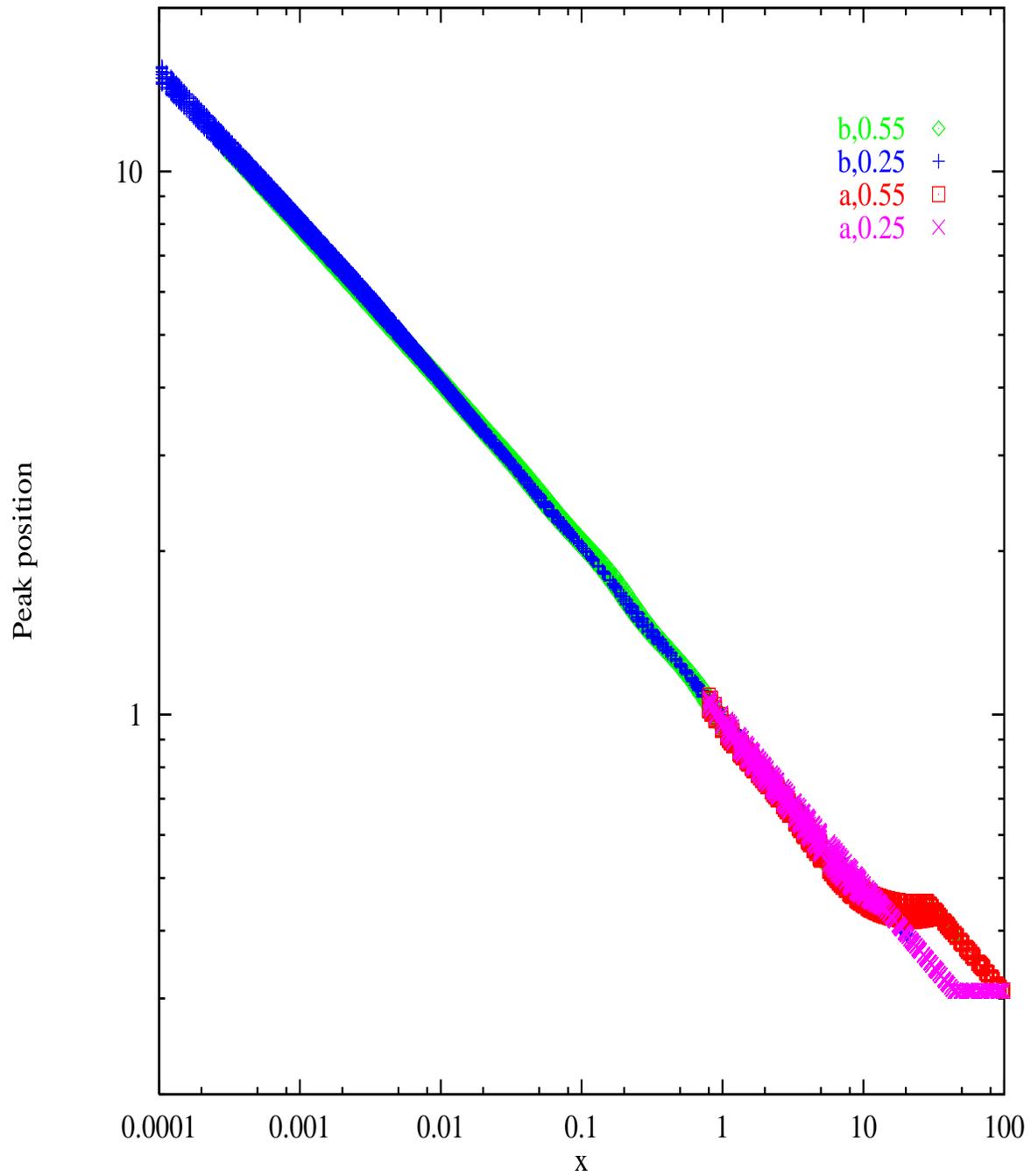,height=18.cm,width=16.cm}
\vspace{-.50cm}
\caption{  Same as Fig. 1 but plotted versus new dynamical variable $x$.}
\end{minipage}
}\end{center}
%\vspace {-0.25cm}
\end{figure}
%%%------------g(k,p,mu,T) -----------------%%%%%%%%
\begin{figure}[!ht]
\begin{center}{
\hspace{-.5cm}
\begin{minipage}{16.cm}
\psfig{figure=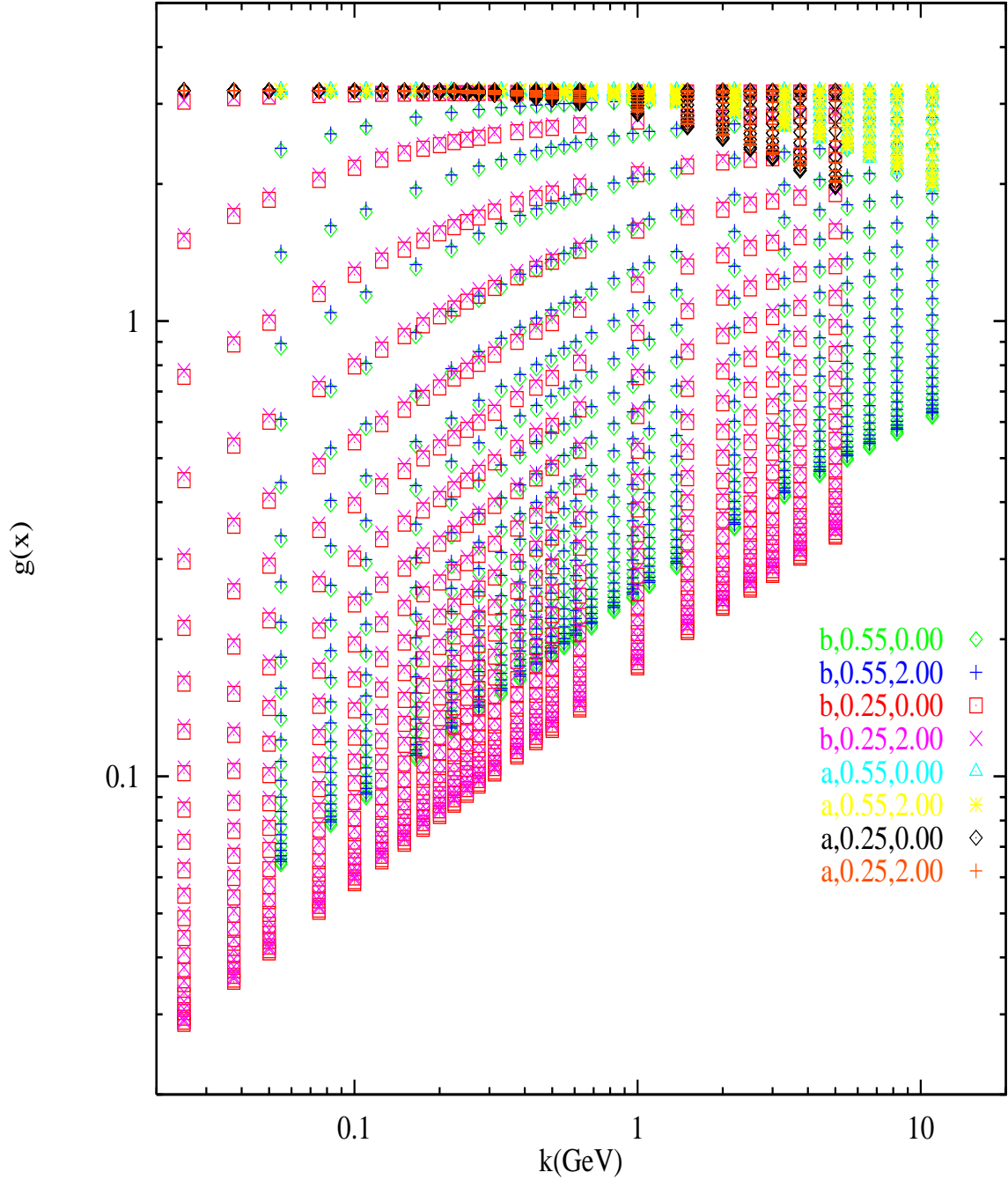,height=18.cm,width=16.0cm}
\vspace{-.50cm}
\caption{  The  new  dimensionelss  emission  function $g(k,p_\parallel
,\kappa ,T)$ mentioned in the text versus photon energy. This data  is
generated from the variational method for bremsstrahlung and ${\bf aws
}$ processes and shown by labels ~$b$~ and ~$a$~ in figure. The labels
in figure show process,temperatures and chemical potential values. For
each  process  at  a  given  temperature  and density values, the data
contains several $p_\parallel $ values all in same symbol and color.}
\end{minipage}
}\end{center}
%\vspace {-0.25cm}
\end{figure}
\begin{figure}[!ht]
\begin{center}{
\hspace{-.5cm}
\begin{minipage}{16.cm}
\psfig{figure=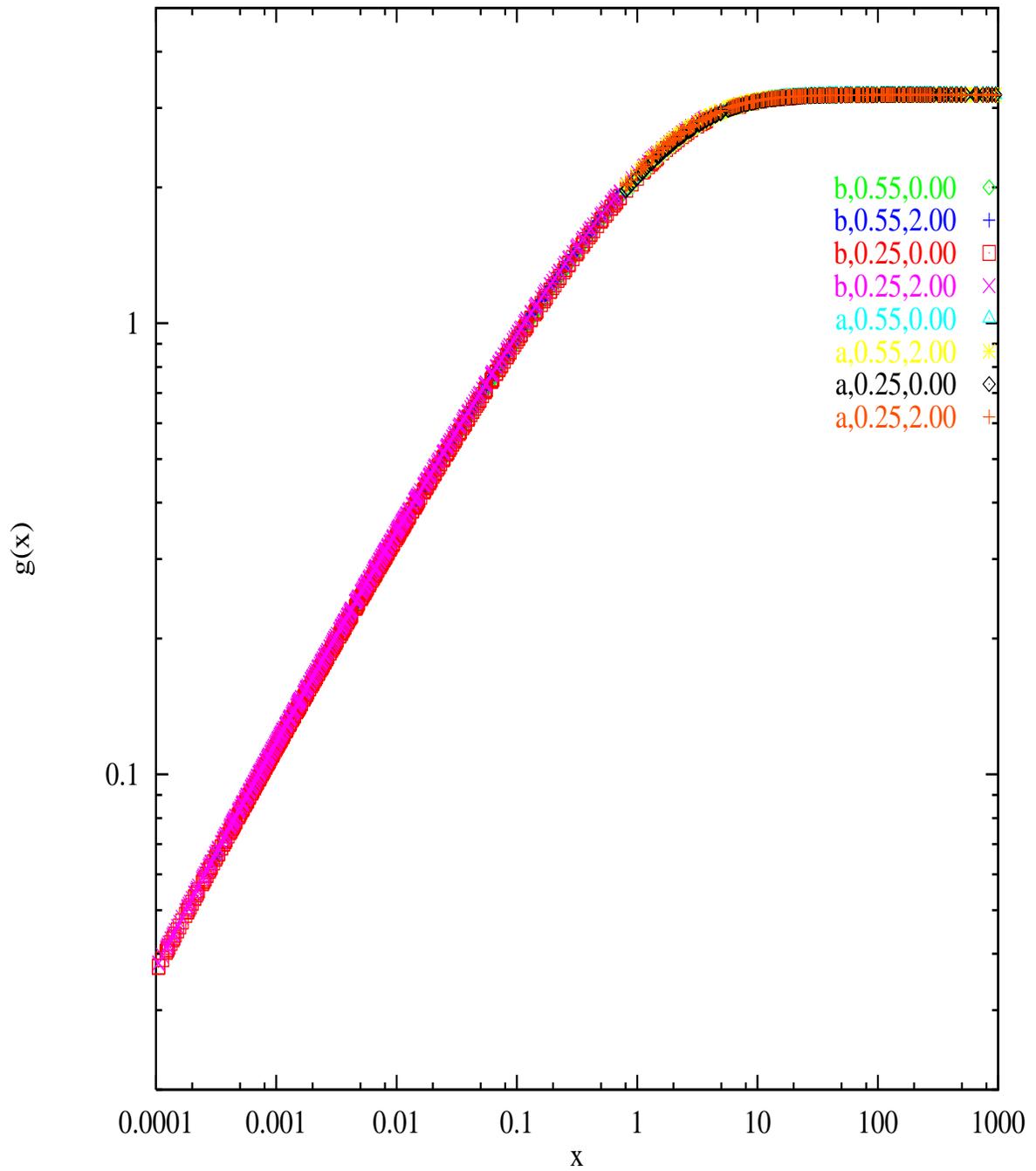,height=18.cm,width=16.0cm}
\vspace{-.50cm}
\caption{  Same  as  Fig.3,  the  new  dimensionelss emission function
$g(x)$ versus dynamical variable $x$. All the cases of Fig. 3  are  in
the  figure  but  may  not  be visible due to superposition of various
symbols.}
\end{minipage}
}\end{center}
%\vspace {-0.25cm}
\end{figure}
%%%------------suppression factors at T=0.25GeV -----------------%%%%%%%%
\begin{figure}[!ht]
\begin{center}{
\hspace{-.5cm}
\begin{minipage}{16.cm}
\psfig{figure=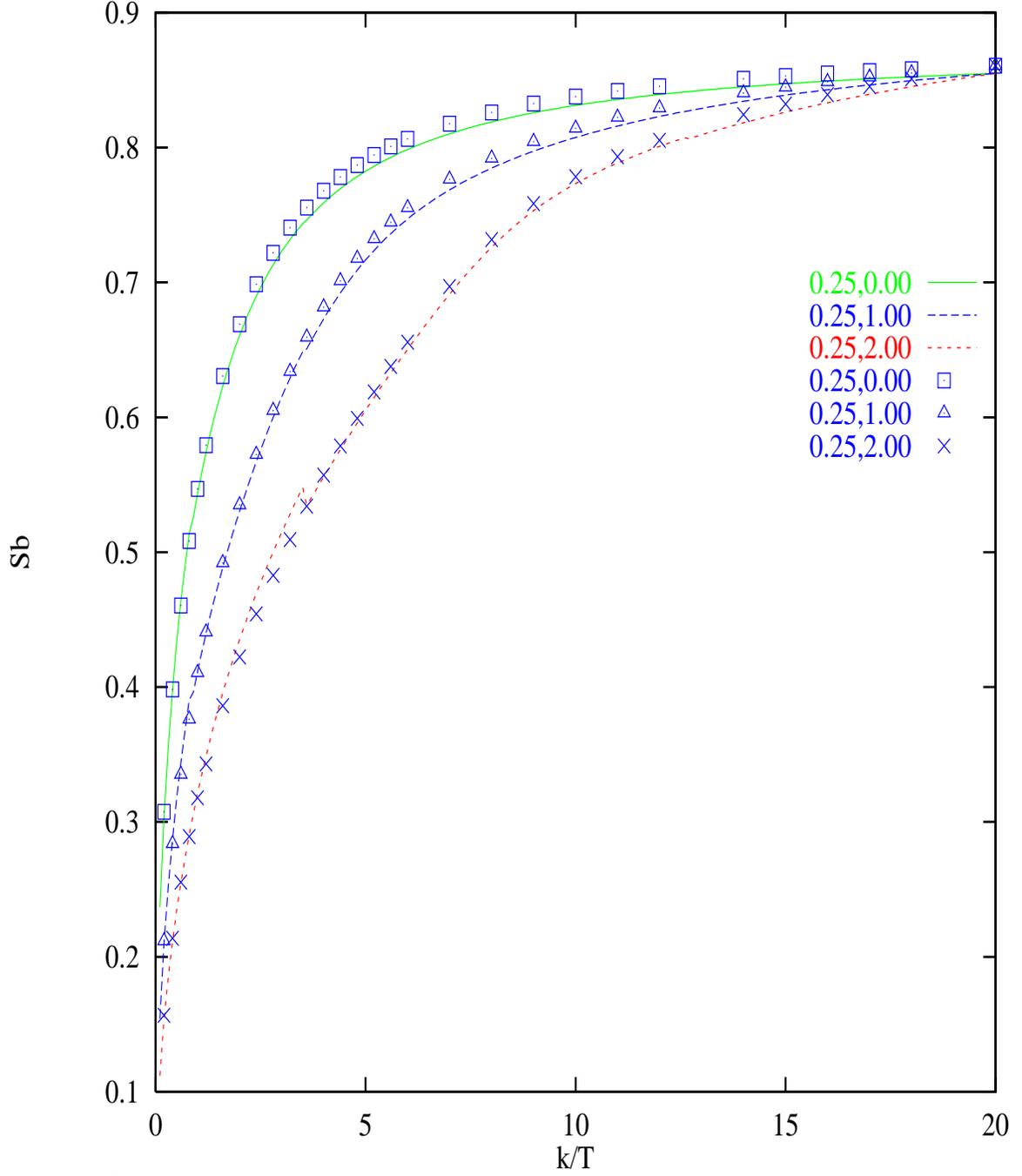,height=18.cm,width=16.0cm}
\vspace{-.50cm}
\caption{ Suppression factors for bremsstrahlung radiation relative to
zero  density  case. The temperature is 0.25GeV and chemical potential
values are mentioned in the figure. The curves are obtained  from  the
empirical  $g(x)$  function, and the symbols are from rigorous results
using variational method with $n_r=8.$}
\end{minipage}
}\end{center}
%\vspace {-0.25cm}
\end{figure}
\begin{figure}[!ht]
\begin{center}{
\hspace{-.5cm}
\begin{minipage}{16.cm}
\psfig{figure=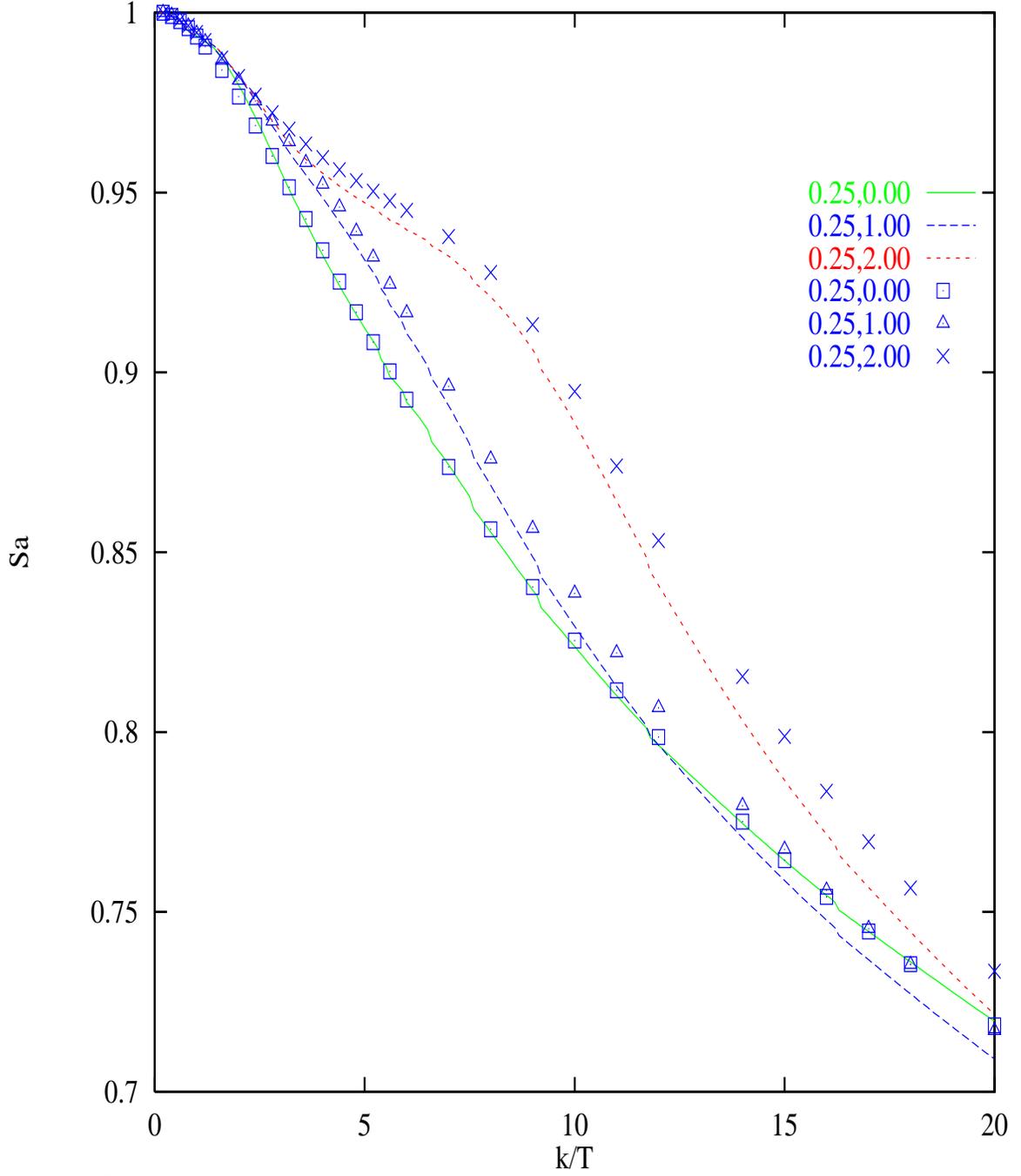,height=18.cm,width=16.0cm}
\vspace{-.50cm}
\caption{ Suppression factors for radiation from $\bf aws$ relative to
zero  density  case.  The  temperature  is  0.25GeV  and  the chemical
potential values are mentioned in the figure. The curves are  obtained
from  the empirical $g(x)$ function, and the symbols are from rigorous
calculations using variational method with $n_r=8.$}
\end{minipage}
}\end{center}
%\vspace {-0.25cm}
\end{figure}
%%%------------suppression factors at T=0.55GeV -----------------%%%%%%%%
\begin{figure}[!ht]
\begin{center}{
\hspace{-.5cm}
\begin{minipage}{16.cm}
\psfig{figure=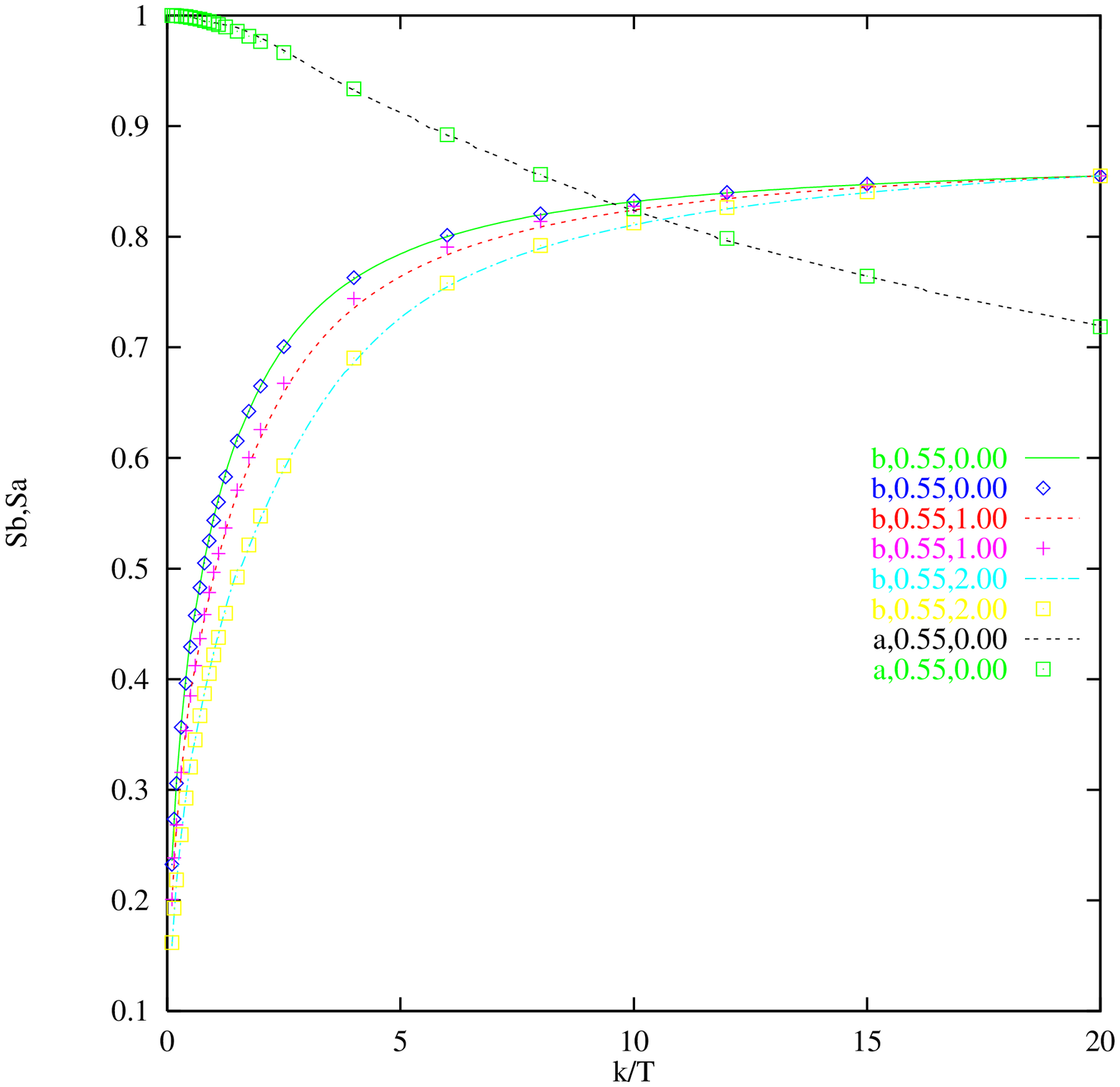,height=18.cm,width=16.0cm}
\vspace{-.50cm}
\caption{ Suppression factors for bremsstrahlung and $\bf aws$ radiation
relative to
zero  density  case. The temperature is 0.55GeV and chemical potential
values are mentioned in the figure. The curves are obtained  from  the
empirical  $g(x)$  function, and the symbols are from rigorous results
using variational method with $n_r=8.$}
\end{minipage}
}\end{center}
%\vspace {-0.25cm}
\end{figure}
%%%------------suppression factors at T=0.55GeV -----------------%%%%%%%%
\begin{figure}[!ht]
\begin{center}{
\hspace{-.5cm}
\begin{minipage}{16.cm}
\psfig{figure=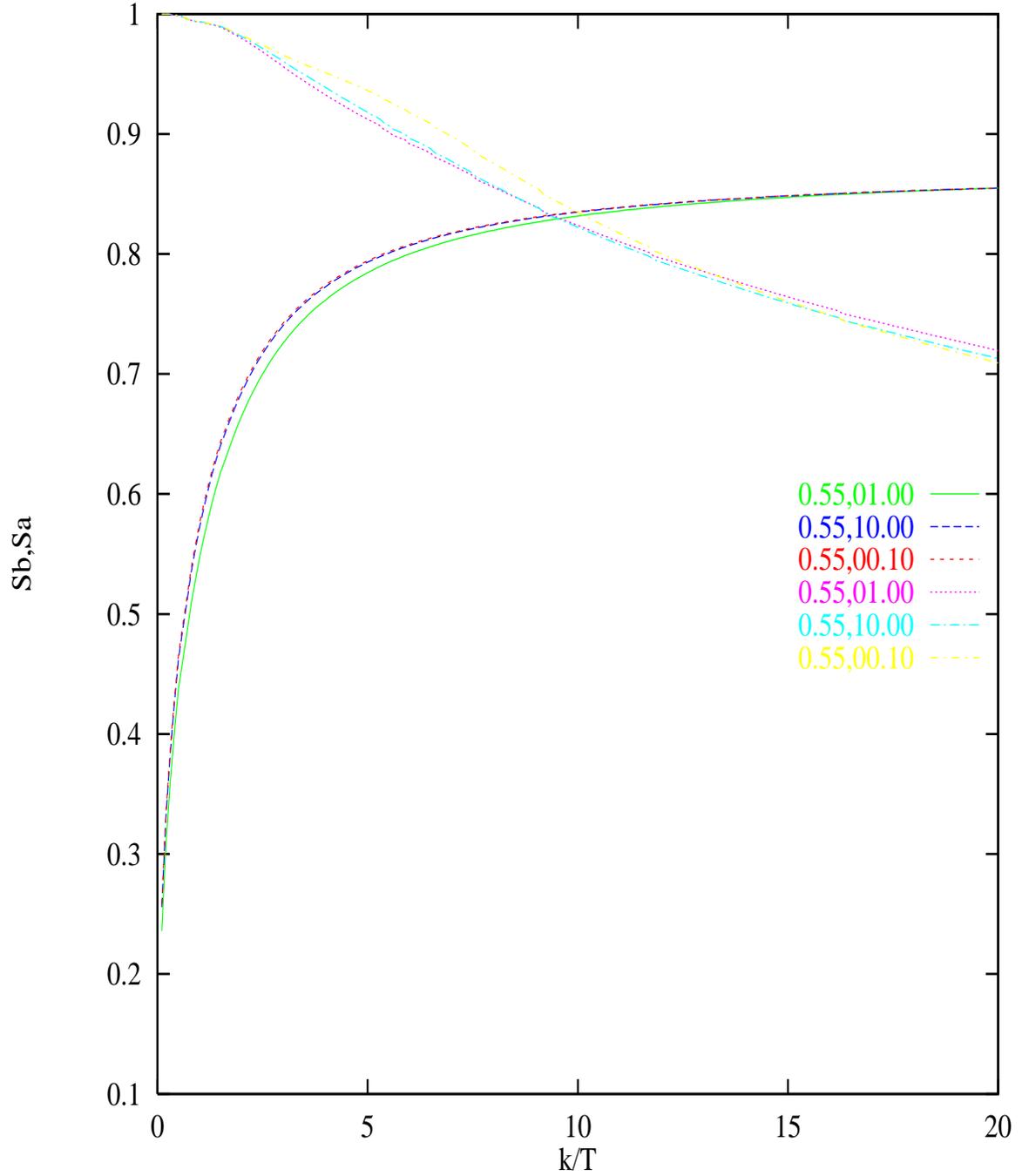,height=18.cm,width=16.0cm}
\vspace{-.50cm}
\caption{ Suppression factors for bremsstrahlung and $\bf aws$ radiation
relative to  saturated plasma, represented by labels ~b~ and ~a~.
The temperature is 0.55GeV and the $\lambda_g/\lambda_q$
values are mentioned in the figure. The curves are obtained  from  the
empirical  $g(x)$  function.}
\end{minipage}
}\end{center}
%\vspace {-0.25cm}
\end{figure}
%%%%%%--------------------------------------------------------------------

%%%------------table of normalisation factors------------------%%%%%%%%
\begin {center}{
\begin{table}
\begin{tabular}
{|c|c|c|c|c|c|c|c|c|c|}
\hline
 $T=0.25$ &  &  &  & &~& $T=0.35$ &  &  &  \\
\hline
 $\mu$ & $\kappa$ & $J_T-J_L$ & $f_b$ & $f_a$ && $\kappa$ & $J_T-J_L$ & $f_b$ & $f_a$ \\
\hline
  0.00& 0.25000& 2.05215& 0.85500& 1.00000&&0.25000& 2.05215& 0.85500& 1.00000\\
  0.25& 0.25589& 2.02793& 0.83813& 0.99791&&0.25311& 2.03926& 0.84595& 0.99888\\
  0.50& 0.26943& 1.97486& 0.80245& 0.99351&&0.26119& 2.00672& 0.82397& 0.99612\\
  0.75& 0.28385& 1.92201& 0.76243& 0.98939&&0.27156& 1.96682& 0.79697& 0.99287\\
  1.00& 0.29572& 1.88105& 0.72491& 0.98637&&0.28190& 1.92892& 0.76926& 0.98991\\
  1.50& 0.31102& 1.83136& 0.65706& 0.98293&&0.29855& 1.87162& 0.71568& 0.98570\\
  2.00& 0.31912& 1.80636& 0.59987& 0.98129&&0.30940& 1.83649& 0.66820& 0.98327\\
\hline
 $T=0.45$ &  &  & & && $T=0.55$ &  &  &  \\
\hline
  0.00& 0.25000& 2.05215& 0.85500& 1.00000&&0.25000& 2.05215& 0.85500& 1.00000\\
  0.25& 0.25191& 2.04421& 0.84940& 0.99931&&0.25129& 2.04678& 0.85122& 0.99953\\
  0.50& 0.25715& 2.02283& 0.83485& 0.99748&&0.25492& 2.03184& 0.84097& 0.99824\\
  0.75& 0.26452& 1.99366& 0.81551& 0.99504&&0.26032& 2.01017& 0.82653& 0.99641\\
  1.00& 0.27274& 1.96239& 0.79439& 0.99252&&0.26673& 1.98514& 0.80995& 0.99434\\
  1.50& 0.28815& 1.90691& 0.75176& 0.98826&&0.28009& 1.93542& 0.77486& 0.99041\\
  2.00& 0.30001& 1.86678& 0.71166& 0.98536&&0.29177& 1.89445& 0.74034& 0.98734\\
\hline
 $T=0.55$ &  &  & & &&  &  &  &  \\
\hline
   1.00& 0.25000& 2.05215& 0.85500& 1.00000&&&&&\\
   0.10& 0.31850& 1.80824& 0.85782& 0.98141&&&&&\\
   10.0& 0.22898& 2.14498& 0.88409& 1.00848&&&&&\\
   5.00& 0.23494& 2.11758& 0.88153& 1.00590&&&&&\\
\hline
\end {tabular}
\caption{Normalization factors as a function of baryon density represented by quark
chemical potential (in GeV) for various temperatures (in GeV). The $\kappa$ value
used in the integral equation and the $J_T,J_L$ values required are  also given in Table.
The last four entries of Table are for unsaturated plasma for the parameters
in text, with the second column for $\lambda_g/\lambda_q$.}
\end{table}
}\end {center}

\begin{thebibliography}{99}
\bibitem{peitz} Thomas Peitzman and Markus H. Thoma,  hep-ph/0111114.
\bibitem{auren1}P. Aurenche, F. Gelis, H. Zaraket  and R. Kobes ,  Phys.
Rev. {\bf D58} 085003 (1998), [hep-ph/9804224]~;\\
P. Aurenche, F. Gelis, R. Kobes  and E. Petitgirard,  Phys.
Rev. {\bf D54} 5274 (1996) [hep-ph/9604398];
Z. Phys. ${\bf C75}$,315 (1996).
\bibitem{auren2}P. Aurenche, F. Gelis, R. Kobes  and H. Zaraket,
Phys. Rev. {\bf D61} 116001 (2000) [hep-ph/9911367]
\bibitem{auren3}P. Aurenche, F. Gelis, and  H.  Zaraket, Phys. Rev. {\bf D62} 096012 (2000) [hep-ph/0003326]
\bibitem{auren4}P. Aurenche, F. Gelis, and  H.  Zaraket,  hep-ph/0204146
\bibitem{arnold1}Peter  Arnold,  Guy  D.  Moore  and  Laurence  G.   Yaffe, JHEP 0112 (2001) 009, [hep-ph/0111107]
\bibitem{arnold2}Peter  Arnold,  Guy  D.  Moore  and  Laurence  G.   Yaffe, JHEP 0111 (2001) 057, [hep-ph/0109064].
\bibitem{arnold3}Peter  Arnold,  Guy  D.  Moore  and  Laurence  G.   Yaffe, hep-ph/0204343.
\bibitem{svs1}S. V. S. Sastry, hep-ph/0208103.
\end{thebibliography}
\end{document}